\documentstyle[12pt,psfig]{article}
\newcommand{\be}{\begin{equation}}
\newcommand{\ee}{\end{equation}}
\newcommand{\bag}{B^{1/4}}
\newcommand{\rgap}{R_{\rm gap}}
\newcommand{\vcrust}{V_{\rm crust}}
\newcommand{\Eos}{Equation of state~}
\newcommand{\eos}{equation of state~}

\newcommand{\ecrusti}{\epsilon_{\rm crust}}
\newcommand{\edrip}{\epsilon_{\rm drip}}
\newcommand{\pdrip}{P_{\rm drip}}
\newcommand{\gcmt}{{\rm g/cm}^3}
\newcommand{\lsim}{\stackrel{\textstyle <}{_\sim}}
\newcommand{\gsim}{\stackrel{\textstyle >}{_\sim}}
\newcommand{\eosp}{equation of state}
\newcommand{\beqn}{\begin{eqnarray}}
\newcommand{\eeqn}{\end{eqnarray}}
\newcommand{\okgr}{\Omega_{\rm K}}
\newcommand{\pkgr}{P_{\rm K}}
\newcommand{\msun}{M_{\odot}}
\newcommand{\msec}{\rm msec}
\newcommand{\mmin}{M_{\rm min}}
\newcommand{\rcore}{R_{\rm core}}
\newcommand{\icrust}{I_{\rm crust}}
\newcommand{\itotal}{I_{\rm total}}
\newcommand{\rdrip}{R_{\rm drip}}
\newcommand{\rsurf}{R_{\rm surf}}
\newcommand{\smh}{strange matter hypothesis~}

 1
\font\elevenrm=cmr10 scaled\magstep 1
 1

\def\refind{\hang\noindent}
\newcommand{\doe}
{This work was supported by the Director, Office of Energy Research,
Office of High Energy and Nuclear Physics, Division of Nuclear Physics, of 
the U.S. Department of Energy under Contract DE-AC03-76SF00098.}
\newcommand{\dateofdoc}{\today}
\newcommand{\lbl}{\begin{flushright} LBNL--39305 \\[8ex] \end{flushright}}

\begin{document}

\begin{titlepage}
\lbl
\begin{center}
\begin{Large}
{\bf From Quark Matter to Strange Machos}
\end{Large}
\end{center}

\vspace{0.7cm}
\begin{center}
\begin{large}
F. Weber$\,^{a,b}$, Ch. Schaab$\,^a$, M. K. Weigel$\,^a$, and N. K. 
Glendenning$\,^b$
\end{large}
\end{center}

\vspace{0.5cm}
\begin{center}
$^a\,$Institute for Theoretical Physics, Ludwig-Maximilians University \\
Theresienstrasse 37, D-80333 Munich, Germany\\[3ex]
$^b\,$Nuclear Science Division and Institute for Nuclear \& Particle
Astrophysics\\
Lawrence Berkeley National Laboratory, MS 70A-3307 \\
University of California, Berkeley, California 94720, USA
\end{center}
\vspace{3cm}

\begin{center}
\dateofdoc \\[22ex]
\end{center}

\begin{quote}
\begin{center} 
{Presented by F. Weber at the \\
Vulcano Workshop 1996 \\
Frontier Objects in Astrophysics and Particle Physics \\
May 27 -- June 1, Vulcano, Italy \\
To be published by the Societa Italiana di Fisica}
\end{center}
\end{quote}
\end{titlepage}

\vspace*{1.8cm}
  \centerline{\bf From Quark Matter to Strange Machos}
\vspace{1cm}
  \centerline{F. Weber$\,^{a,b}$, Ch. Schaab$\,^a$, M. K. Weigel$\,^a$, 
              and N. K. Glendenning$\,^b$}
\vspace{1.4cm}
\centerline{$^a\,$Institute for Theoretical Physics, Ludwig-Maximilians
  University}
\centerline{\elevenrm Theresienstrasse 37, D-80333 Munich, Germany} 
\centerline{$^b\,$Nuclear Science Division and Institute for Nuclear \& 
            Particle  Astrophysics}
\centerline{\elevenrm Lawrence Berkeley National Laboratory, MS 70A-3307}
\centerline{\elevenrm University of California, Berkeley, California 94720, 
USA}
\vspace{3cm}
\begin{abstract} 
  This paper gives an overview of the properties of all possible equilibrium
  sequences of compact strange-matter stars with nuclear crusts, which range
  from strange stars to strange dwarfs. In contrast to their non-strange
  counterparts, --neutron stars and white dwarfs--, their properties are
  determined by two (rather than one) parameters, the central star density and
  the density at the base of the nuclear crust. This leads to stellar
  strange-matter configurations whose properties are much more complex than
  those of the conventional sequence.  As an example, two generically different
  categories of stable strange dwarfs are found, which could be the observed
  white dwarfs.  Furthermore we find very-low-mass strange stellar objects,
  with masses as small as those of Jupiter or even lighter planets.  Such
  objects, if abundant enough in our Galaxy, should be seen by the presently
  performed gravitational microlensing searches.
\end{abstract}
\vspace{2.0cm}

\goodbreak
\section{Introduction}\label{sec:intro}

The theoretical possibility that strange quark matter may be absolutely stable
with respect to iron, that is, the energy per baryon is below 930 MeV, has been
pointed out by Bodmer\ (1971), Terazawa\ (1979), and Witten\ (1984).  This
so-called strange matter hypothesis constitutes one of the most startling
possibilities of the behavior of superdense nuclear matter, which, if true,
would have implications of fundamental importance for cosmology, the early
universe, its
\begin{figure}[tb]
\begin{center}
\leavevmode
\mbox{\psfig{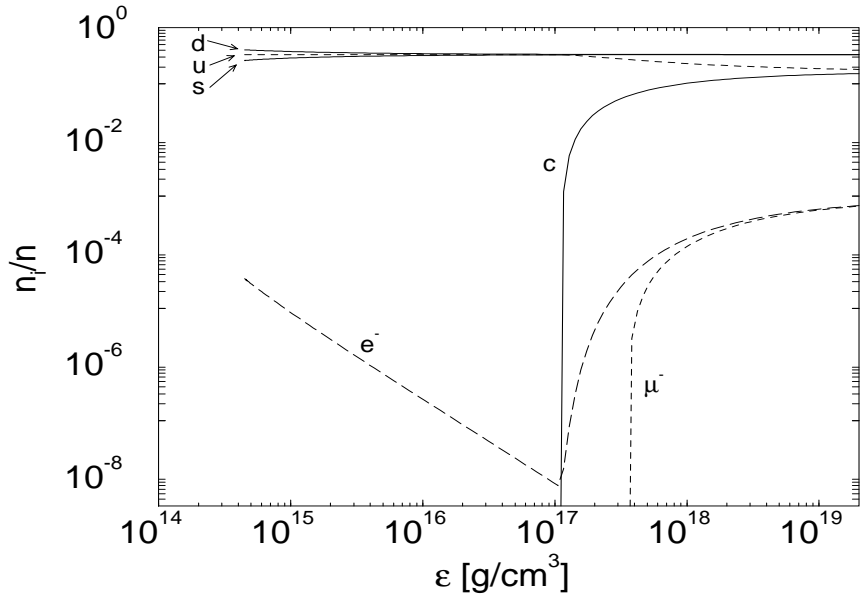}}
\caption[Relative densities of quarks and leptons in cold, beta-stable, 
electrically charge neutral quark matter as a function of mass density.]  {\em
  {Relative densities of quarks and leptons, $n_i/n$, where $n$ denotes the
    total quark density, in cold, beta-stable, electrically charge neutral
    quark-star matter as a function of energy density ($\bag=145$ MeV)
    (Kettner et al,\ 1995a).}}
\label{fig:1.5}
\end{center}
\end{figure} evolution to the present day, astrophysical compact objects, and
laboratory physics (for an overview, see Madsen and Haensel\ (1991), and 
Table\ 1).  Even to the present day there is no sound
\begin{figure}[tb]
\begin{center}
\leavevmode
\mbox{\psfig{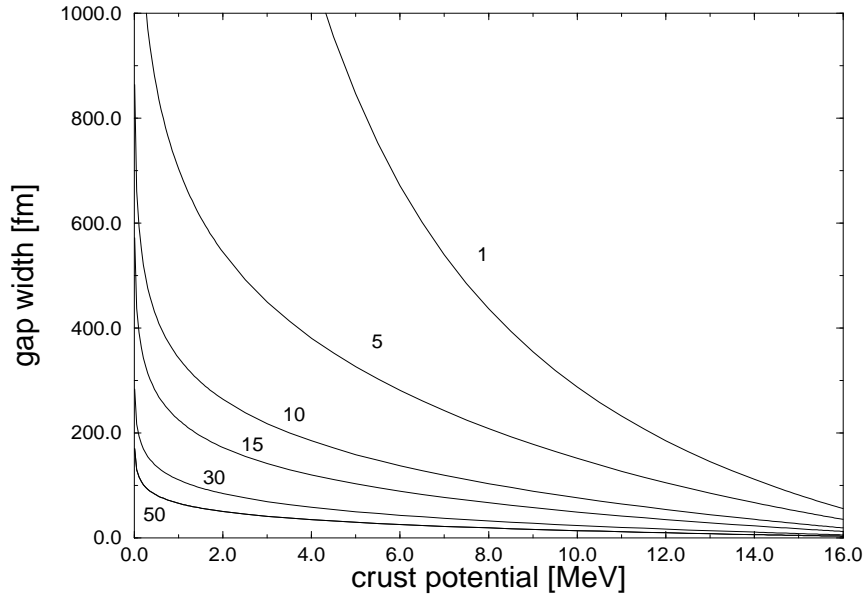}}
\caption[Gap width versus electrostatic crust potential, for different 
temperatures]{\em {Gap width, $\rgap$, versus electrostatic crust potential,
    $e\vcrust$. The labels refer to temperature (in MeV).}}
\label{fig:vvsgap}
\end{center}
\end{figure}
scientific basis on which one can either confirm or reject the
hypothesis, so that it remains a serious possibility of fundamental
significance for various phenomena.
\begin{table}[tb]
\begin{center}
\begin{tabular}{|l|l|} \hline
Phenomenon                      &References \\ \hline 
Centauro cosmic ray events      &Chin et al.\ (1979), Bjorken et al.\ (1979),\\
                                &Witten\ (1984)\\
High-energy gamma ray sources   &Jaffe\ (1977), Baym et al.\ (1985) \\  
Strange matter hitting the earth:             &                   \\
~~~~strange meteors                 &De R{\'{u}}jula et al.\ (1984) \\
~~~~nuclearite-induced earthquakes  &De R{\'{u}}jula et al.\ (1984) \\
~~~~strange nuggets in cosmic rays  &Terazawa\ (1991,1993)  \\
Strange matter in supernovae       &Michel\ (1988), Benvenuto et al.\ (1989),\\
                                    &Horvath et al.\ (1992) \\
Strange star (pulsar) phenomenology &Alcock et al.\ (1986), Haensel et al.\
                                     (1986), \\
                 &Alcock et al.\ (1988), Glendenning\ (1990), \\
                 &Glendenning et al.\ (1992)\\
Strange dwarfs   &Glendenning et al.\ (1995a),\\
                 &Glendenning et al.\ (1995b) \\
Strange planets  &Glendenning et al.\ (1995a),\\
                 &Glendenning et al.\ (1995b) \\
Burning of neutron stars to strange stars   
                 &Olinto\ (1987), Horvath et al.\ (1988),\\
                 &Frieman et al.\ (1989)\\
Gamma-ray bursts    &Alcock et al.\ (1986), Horvath et al.\ (1993)\\
Cosmological aspects of strange matter 
                    &Witten\ (1984), Madsen et al.\ (1986),\\
                    &Madsen\ (1988), Alcock et al.\ (1988) \\
Strange matter as compact energy source         &Shaw et al.\ (1989) \\
Strangelets in nuclear collisions  
               &Liu et al.\ (1984), Greiner et al.\ (1987),\\
               &Greiner et al.\ (1988)\\
\hline
\end{tabular}
\caption[Strange Matter Phenomenology]{\em {Overview of strange matter 
    phenomenology}}\label{tab:over}
\end{center}
\end{table}

On theoretical scale arguments, strange quark matter is as plausible a
\begin{figure}[tb]
\begin{center}
\leavevmode
\mbox{\psfig{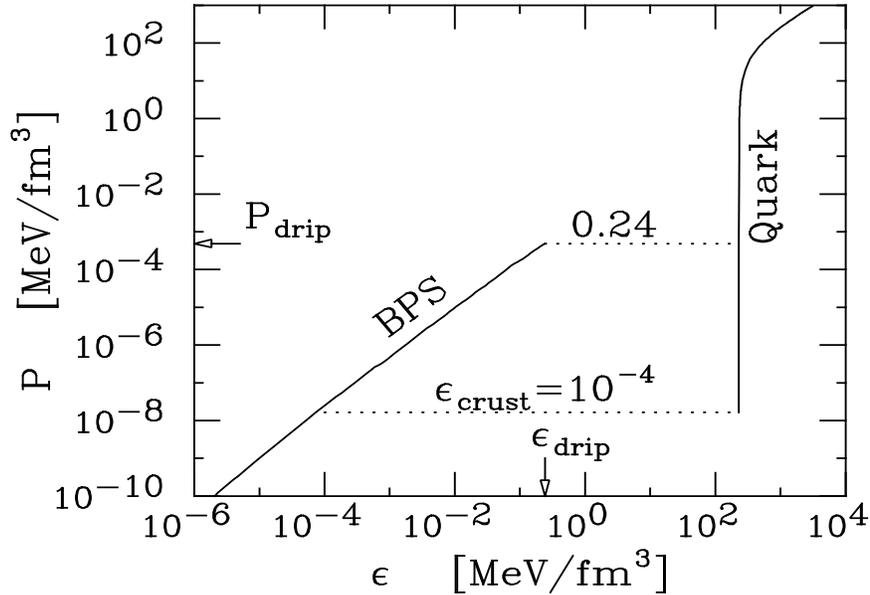}}
\caption[\Eos of a strange star surrounded by a nuclear crust with inner
crust density lower than neutron drip, $\ecrusti \leq \edrip$.]{\em {\Eos of a
    strange star surrounded by a nuclear crust.  $\pdrip(\edrip)$ denotes the
    pressure at the maximum possible inner crust density determined by neutron
    drip, $\ecrusti = 0.24~MeV/fm^3$. Any inner crust value smaller than that
    is possible. As an example, we show the \eos for $\ecrusti =10^{-4}~
    MeV/fm^3$.}}
\label{fig:eos}
\end{center}
\end{figure} ground state as the confined state of hadrons (Witten, 1984; Farhi
and Jaffe, 1984; Glendenning, 1990) Unfortunately it seems unlikely that QCD
calculations will be accurate enough in the
\begin{table}[tb]
\begin{center}
\begin{tabular}{|l|l|} \hline
Experiment                                   &References         \\ \hline
Cosmic ray searches for strange nuggets:     &                              \\
~~~~~balloon-borne experiments               &Saito\ (1990, 1995)            \\
~~~~~MACRO                                   &MACRO\ (1992)                  \\
~~~~~IMB                                     &De R{\'{u}}jula et al.\ (1983) \\
~~~~~tracks in ancient mica                  &De R{\'{u}}jula et al.\ (1984) \\
                                             &Price\ (1984) \\ 
Rutherford backscattering of $^{238}$U and $^{208}$Pb 
                                             &Br{\"{u}}gger et al.\ (1989)  \\
Heavy-ion experiments at BNL: E864, E878, E882-B, &Thomas et al.\ (1995) \\
$~$~~~~~~~~~~~~~~~~~~~~~~~~~~~~~~~~~~~~~~~~~~E886, E-888, E896-A  &      \\
Heavy-ion experiments at CERN: NA52               &Thomas et al.\ (1995) \\
\hline
\end{tabular}
\caption[Search experiments for strange matter]
{\em {Overview of search experiments for strange matter}}\label{tab:labexp}
\end{center}
\end{table} foreseeable future to give a definitive prediction on the stability
of strange matter, and one is left with experiment, Table\ 2, and astrophysical
tests, as performed here, to either confirm or reject the hypothesis.

One striking implication of the hypothesis would be that pulsars, which are
conventionally interpreted as rotating neutron stars, almost certainly would be
rotating strange stars (strange pulsars) (Witten, 1984; Haensel, Zdunik, and
Schaeffer,\ 1986; Alcock, Farhi, and Olinto,\ 1986; Glendenning,\ 1990).  Part
of this paper deals with an investigation of the properties of such objects.
In addition to this, we develop the complete sequence of strange stars with
nuclear crusts, which ranges from the compact members, with properties similar
to those of neutron stars, to white dwarf-like objects (strange dwarfs), to
planetary-like strange matter objects, and discuss their
\begin{figure}[tb]
\begin{center}
\leavevmode
\mbox{\psfig{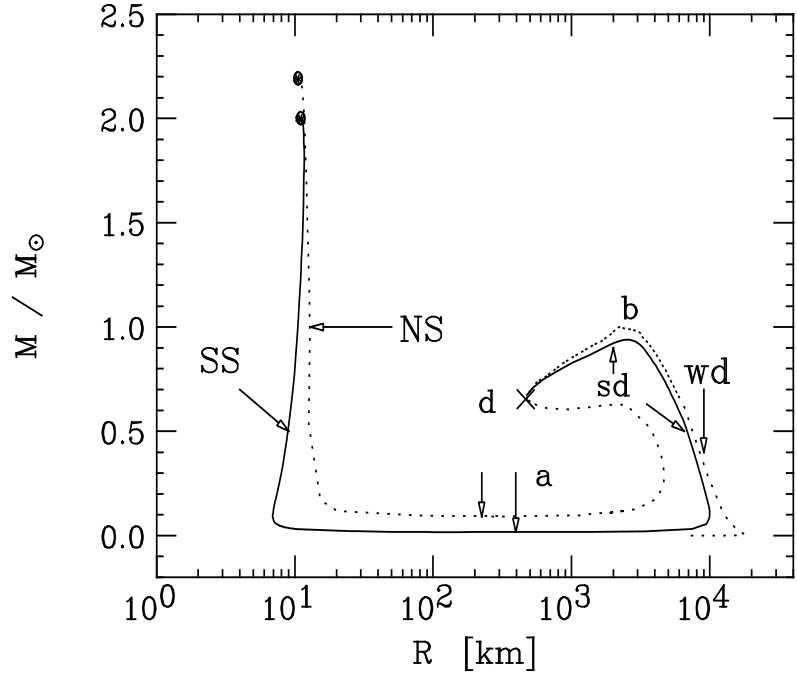}}
\caption[Mass versus radius of strange-star configurations with nuclear crust.]
{\em {Mass versus radius of strange-star configurations with nuclear crusts
    (solid curve) and gravitationally bound stars (dotted).  (NS=neutron star,
    SS=strange star, wd=white dwarf, sd=strange dwarf.) The cross denotes the
    termination point of the strange-dwarf sequence ($\ecrusti=\edrip$). The
    dots and vertical arrows refer to the maximum- and minimum-mass star of
    each sequence.}}
\label{fig:1}
\end{center}
\end{figure} stability against acoustical vibrations (Glendenning, Kettner, and
Weber,\ 1995a,b; Kettner, Weber, Weigel, and Glendenning,\ 1995b).  The
properties with respect to which strange-matter stars differ from their
non-strange counterparts are discussed, and observable signatures of strange
stars are pointed out.

\goodbreak
\section{Quark-lepton composition of strange matter}\label{sec:qlc}

The relative quark/lepton composition of quark-star matter at zero temperature
is shown in Fig.\ 1.  All quark flavor states that become populated at the
densities shown are taken into account. (Strange and charm quark masses of
respectively 0.15 GeV and 1.2 GeV are assumed.) Since stars in their lowest
energy state are electrically charge neutral to very high precision
(Glendenning, 1985), any net positive quark charge must be balanced by leptons.
In general, as can be seen in Fig.\ 1, there is only little need for leptons,
since charge neutrality can be achieved essentially among the quarks
themselves.  The concentration of electrons is largest at the lower densities
of Fig.\ 1 due to the finite $s$-quark mass which leads to a deficit of net
negative quark charge, and at densities beyond which the $c$-quark state
becomes populated which increases the net positive quark charge.

\goodbreak
\section{Nuclear crusts on strange stars}\label{sec:ncss}

The presence of electrons in strange quark matter is crucial for the possible
existence of a nuclear crust on such objects.  As shown by Alcock, Farhi, and
Olinto (1996), and Kettner, Weber, Weigel, and Glendenning (1995a), the
electrons, because they are bound to strange matter by the Coulomb force rather
than the strong force, extend several hundred fermi beyond the surface of the
strange star.  Associated with this electron displacement is a electric dipole
layer which can support, out of contact with the surface of the strange star, a
crust of nuclear material, which it polarizes (Alcock, Farhi, and Olinto,\ 
1986).  The maximal possible density at the base of the crust (inner crust
density) is determined by neutron drip, which occurs at about $4.3\times
10^{11}~\gcmt$.  (Free neutrons in the star cannot exist.  These would be
dissolved into quark matter as they gravitate into the strange core.  Therefore
the maximum density of the crust is strictly limited by neutron drip.)

The determination of the electrostatic electron potential at the surface of a
strange star performed by Alcock, Farhi, and Olinto (1986) has been extended to
finite temperatures only recently (Kettner, Weber, Weigel, and Glendenning,\ 
1995a).  The results obtained there for the gap between the surface of the
star's strange core and the base of the inner crust are shown in Fig.\ 2.  A
minimum value of $\rgap\sim 200$ fm was established by Alcock, Farhi, and
Olinto (1986) as the lower bound on $\rgap$ necessary to guarantee the crust's
security against strong interactions with the strange-matter core.  For this
value one finds from Fig.\ 2 that a hot strange pulsar with $T\sim 30$ MeV can
only carry nuclear crusts whose electrostatic potential at the base is rather
smaller, $e\vcrust\lsim 0.1$ MeV.  Crust potentials in
\begin{table}[tbh]
\begin{center}
\begin{tabular}{|l|l|l|} \hline
Features  of strange quark-matter stars ~~~  &observable~~ &definite signal \\
\hline \hline 
$\bullet$ Strange Stars:                   &                    &          \\
Small rotational periods, $P<1$ msec       &yes$\,^\dagger$     &possibly  \\ 
$\bullet$ Light, planetary-like objects    &yes                 &no        \\
$\bullet$ Strange Dwarfs (white-dwarf-like) &yes       &to be studied$\,^*$   
\\
$\bullet$ Cooling behavior                 &yes        &possibly \\
$\bullet$ Glitches                         &yes        &to be studied$\,^*$ \\
$\bullet$ Post-glitch behavior             &yes        &to be studied$\,^*$\\
\hline 
\end{tabular}
\caption[Features of strange quark-matter stars]{\em {Features of strange
    quark-matter stars.($\;^\dagger$Until recently rotational periods of $P\sim
    1$ millisecond were the borderline of detectability.~$^*$Presently under
    investigation.)}}\label{tab:feat}
\end{center}
\end{table} the range of $8$--$12$ MeV, which are expected for a crust at
neutron drip density (Alcock, Farhi, and Olinto,\ 1986), are only possible for
core temperatures of $T\lsim 5$ MeV.  Therefore we conclude that only strange
stars with rather low temperatures (on the nuclear scale) can carry the densest
possible crusts.

\goodbreak
\section{Equation of state of strange stars with crust}\label{sec:eos}

The somewhat complicated situation of the structure of a strange star with
crust described above can be represented by a proper choice of \eos
(Glendenning and Weber,\ 1992), which consists of two parts (Fig.\ 3). At
densities below neutron drip it is represented by the low-density \eos of
charge-neutral nuclear matter, for which we use the Baym-Pethick-Sutherland
\eosp.  The star's strange-matter core is described by the bag model (Freedman
and McLerran,\ 1977; Farhi and Jaffe,\ 1984; Glendenning and Weber,\ 1992;
Kettner, Weber, Weigel, and Glendenning,\ 1995a).

\goodbreak
\section{Properties of strange-matter stars}\label{sec:psec}

\goodbreak
\subsection{Complete sequences of strange-satter stars}

Since the nuclear crusts surrounding the cores of strange stars are bound by
the gravitational force rather than confinement, the mass-radius relationship
of strange-matter stars with crusts is qualitatively similar to the one of
purely gravitationally bound stars -- i.e., neutron stars and white dwarfs --
as illustrated in Fig.\ 4.  The strange-star sequence is computed for the
maximal possible inner crust density, $\ecrusti=\edrip$.  Of course there are
other possible sequences of strange stars with any smaller value of inner crust
density.  Their properties were discussed by Glendenning, Kettner and Weber
(1995a,b).  From the maximum-mass star (dot), the central density decreases
monotonically through the sequence in each case.  The neutron-star sequence is
computed for a representative model for the \eos of neutron star matter, the
relativistic Hartree-Fock \eos (HFV of Weber and Weigel, 1989), which has been
combined at subnuclear densities with the Baym-Pethick-Sutherland \eosp.  Hence
the white dwarfs shown in Fig.\ 4 are computed for the latter.  (For an
overview of the bulk properties of neutron stars, constructed for a
representative collection of modern nuclear equations of state, we refer to
Weber and Glendenning (1992, 1993a,b).)  Those gravitationally bound stars with
radii $\lsim 200$ km and $\gsim 3000$ km represent stable neutron stars and
white dwarfs, respectively.  The fact that strange stars with crust possess
smaller radii than neutron stars leads to smaller rotational mass shedding
(Kepler) periods $\pkgr$, as indicated by the classical expression
$\pkgr=2\pi\sqrt{R^3/M}$. (We recall that mass shedding sets an absolute limit
on rapid rotation.)  Of course the general relativistic expression for $\pkgr$,
given by (Glendenning and Weber,\ 1992; Glendenning, Kettner, and Weber,\ 
1995a) 
\beqn 
\pkgr \equiv \frac{2\, \pi}{\okgr} \, , ~{\rm with} ~~ \okgr =
\omega +\frac{\omega^\prime}{2\psi^\prime} + e^{\nu -\psi} \sqrt{
\frac{\nu^\prime}{\psi^\prime} + \Bigl(\frac{\omega^\prime}{2
\psi^\prime}e^{\psi-\nu}\Bigr)^2} \; ,
\label{eq:okgr}
\eeqn which is to be applied to neutron and strange stars, is considerably more
complicated. However the qualitative dependence of $\pkgr$ on mass and radius
remains valid (Glendenning and Weber,\ 1994).  So one finds that, due to the
smaller radii of strange stars, the complete sequence of such objects (and not
just those close to the mass peak, as is the case for neutron stars) can
sustain extremely rapid rotation (Glendenning, Kettner, and Weber,\ 1995a).  In
particular, a strange star with a typical pulsar mass of $\sim 1.45\,\msun$ can
rotate at (general relativistic) Kepler periods as small as $P \simeq
0.5~\msec$, depending on crust thickness and bag constant (Glendenning and
Weber,\ 1992; Glendenning, Kettner, and Weber,\ 1995a).  This is to be compared
with $\pkgr\sim 1~\msec$ obtained for neutron stars of the same mass (Weber and
Glendenning,\ 1993a,b).

The minimum-mass configuration of the strange-star sequence (labeled `a' in
Fig.\ 4) has a mass of about $\mmin \sim 0.017\, \msun$ (about 17 Jupiter
masses). More than that, we find stable strange-matter stars that can be even
by orders of magnitude lighter than this star, depending on the chosen value of
inner crust density (Glendenning, Kettner, and Weber,\ 1995a,b).  If abundant
enough in our Galaxy, such low-mass strange stars, whose masses and radii
resemble those of ordinary planets (hence one may call such objects strange
planets, or strange MACHOS) could be seen by the gravitational microlensing
searches that are being performed presently.  Strange stars located to the
right of `a' consist of small strange cores ($\rcore\lsim 3$ km) surrounded by
a thick nuclear crust (made up of white dwarf material).  We thus call such
objects strange dwarfs.  Their cores have shrunk to zero at `d'. What is left
is a ordinary white dwarf with a central density equal to the inner crust
density of the former strange dwarf (Glendenning, Kettner, and Weber,\ 
1995a,b).  A detailed stability analysis of strange stars against radial
oscillations (Kettner, Weber, Weigel, and Glendenning,\ 1995a,b) shows that the
strange dwarfs between `b' and `d' in Fig.\ 4 are unstable against the
fundamental eigenmode.  Hence such objects cannot exist stably in nature.
However all other stars of this sequence ($\ecrusti=\edrip$) are stable against
oscillations.  So, in contrast to neutron stars and white dwarfs, the branches
of strange stars and strange dwarfs are stably connected with each other
(Glendenning, Kettner, and Weber,\ 1995a,b).  So far our discussion was
restricted to inner crust densities equal to neutron drip.  For the case
$\ecrusti<\edrip$, we refer to Glendenning, Kettner, and Weber\ (1995a).

\subsection{Glitch behavior of strange pulsars}
\label{ssec:glitch}

A crucial astrophysical test, which the strange-quark-matter hypothesis must
pass in order to be viable, is whether strange quark stars can give rise to the
observed phenomena of pulsar glitches. In the crust quake model an oblate solid
nuclear crust in its present shape slowly comes out of equilibrium with the
forces acting on it as the rotational period changes, and fractures when the
built-up stress exceeds the sheer strength of the crust material. The period
and rate of change of period slowly heal to the trend preceding the glitch as
the coupling between crust and core re-establish their co-rotation.

The only existing investigation which deals with the calculation of the
thickness, mass and moment of inertia of the nuclear solid crust that can exist
on the surface of a rotating, general relativistic strange quark star has been
performed by Glendenning and Weber\ (1992).  Their calculated mass-radius
relationship for strange stars with a nuclear crust, whose maximum density is
the neutron drip density, is shown in Fig.\ 5.
\begin{figure}[tb]
\begin{center}
\parbox[t]{6.5cm}
{\leavevmode \mbox{\psfig{figure=rm.bb,width=6.0cm,height=7.0cm,angle=90}}
  {\caption[Radius as a function of mass of a strange star with crust, and
    radius of the strange star core for inner crust density equal to neutron
    drip, for non-rotating stars. The bag constant is $\bag=160$ MeV. The solid
    dots refer to the maximum-mass model of the sequence]{\em {Radius as a
        function of mass of a non-rotating strange star with crust (Glendenning
        et al,\ 1992).}}\label{fig:radss}}} \ \hskip1.4cm \ 
\parbox[t]{6.5cm}
{\leavevmode \mbox{\psfig{figure=icit.bb,width=6.0cm,height=7.0cm,angle=90}}
  {\caption[The ratio $\icrust/\itotal$ as a function of star mass. Rotational
    frequencies are shown as a fraction of the Kepler frequency. The solid dots
    refer to the maximum-mass models. The bag constant is $\bag=160$ MeV]{\em
      {The ratio $\icrust/\itotal$ as a function of star mass.  Rotational
        frequencies are shown as a fraction of the Kepler frequency, $\okgr$
        (Glendenning et al,\ 1992).}}
\label{fig:cm160}}}
\end{center}
\end{figure} The radius of the strange quark core, denoted $\rdrip$, is shown
by the dashed line, $\rsurf$ displays the star's surface.  (A value for the bag
constant of $\bag=160$ MeV for which 3-flavor strange matter is absolutely
stable has been chosen. This choice represents weakly bound strange matter with
an energy per baryon $\sim 920$ MeV, and thus corresponds to strange quark
matter being absolutely bound with respect to $^{56}{\rm Fe}$).  The radius of
the strange quark core is proportional to $M^{1/3}$ which is typical for
self-bound objects. This proportionality is only modified near that stellar
mass where gravity terminates the stable sequence.

The moment of inertia of the hadronic crust, $\icrust$, that can be carried by
a strange star as a function of star mass for a sample of rotational
frequencies of $\Omega=\okgr,\okgr/2$ and 0 is shown in Fig.\ 6. Because of the
relatively small crust mass of the maximum-mass models of each sequence, the
ratio $\icrust/\itotal$ is smallest for them (solid dots in Fig.\ 6). The less
massive the strange star the larger its radius (Fig.\ 5) and therefore the
larger both $\icrust$ as well as $\itotal$. The dependence of $\icrust$ and
$\itotal$ on $M$ is such that their ratio $\icrust/\itotal$ is a monotonically
decreasing function of $M$.  One sees that there is only a slight difference
between $\icrust$ for $\Omega=0$ and $\Omega=\okgr/2$.

Of considerable relevance for the question of whether strange stars can 
exhibit glitches in rotation frequency, one sees that $\icrust/\itotal$ 
varies between $10^{-3}$ and $\sim 10^{-5}$ at the maximum mass.
If the angular momentum of the pulsar is conserved in the quake
then the relative frequency change and moment of inertia change are equal,
and one arrives at (Glendenning and Weber,\ 1992)
\beqn
       {{\Delta \Omega}\over{\Omega}} \; = \; 
       {{|\Delta I|}\over {I_0}} \; > \;
       {{|\Delta I|}\over {I}} \; \equiv \; f \;
       {\icrust\over I}\; \sim \; (10^{-5} - 10^{-3})\, f \; , 
       ~{\rm with}  \quad 0 < f < 1\; .
\label{eq:delomeg}
\eeqn Here $I_0$ denotes the moment of inertia of that part of the star whose
frequency is changed in the quake. It might be that of the crust only, or some
fraction, or all of the star. The factor $f$ in Eq.\ (2) represents the
fraction of the crustal moment of inertia that is altered in the quake, i.e.,
$f \equiv |\Delta I|/ \icrust$.  Since the observed glitches have relative
frequency changes $\Delta \Omega/\Omega = (10^{-9} - 10^{-6})$, a change in the
crustal moment of inertia of $f\lsim 0.1$ would cause a giant glitch even in
\begin{figure}[tb] 
\begin{center}
\leavevmode
\mbox{\psfig{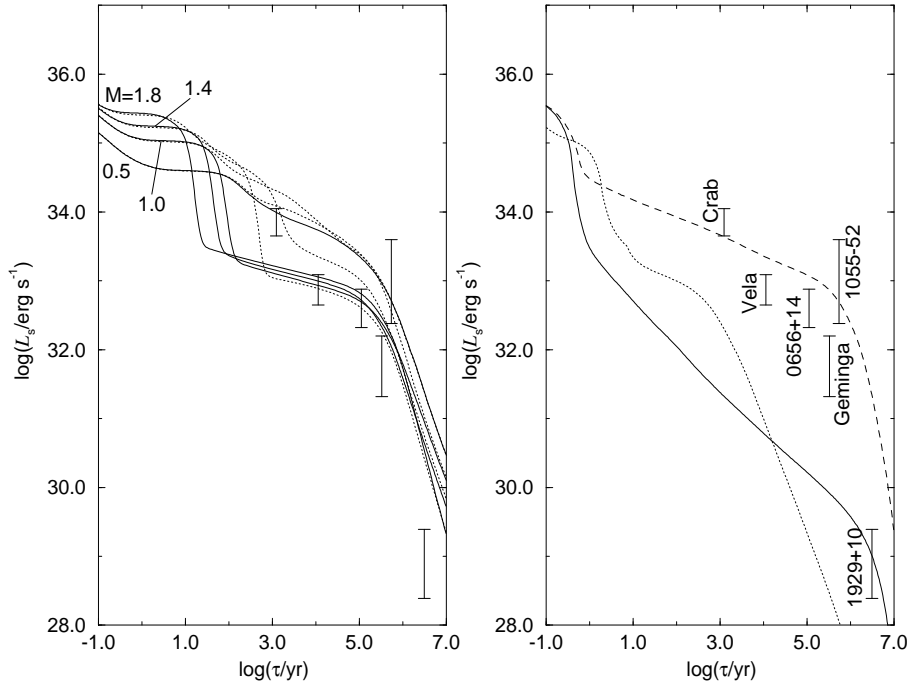}}
  \caption{\em {Left panel: Cooling of neutron stars with pion (solid
      curves) or kaon condensates (dotted curve). Right panel: Cooling of
      $M=1.8\,M_\odot$ strange stars with crust. The cooling curves of lighter
      strange stars, e.g. $M\gsim 1\,M_\odot$, differ only insignificantly from
      those shown here. Three different assumptions about a possible superfluid
      behavior of strange quark matter are made: no superfluidity (solid),
      superfluidity of all three flavors (dotted), and superfluidity of up and
      down flavors only (dashed). The vertical bars denote luminosities of
      observed pulsars.
  \label{fig:cool}}}
\end{center}
\end{figure} 
the least favorable case (for more details, see Glendenning and Weber,\ 1992).
Moreover, we find that the observed range of the fractional change in the
spin-down rate, $\dot \Omega$, is consistent with the crust having the small
moment of inertia calculated and the quake involving only a small fraction $f$
of that, just as in Eq.\ (2).  For this purpose we write (Glendenning and
Weber,\ 1992) \beqn { {\Delta \dot\Omega}\over{\dot\Omega } } \; = \; { {\Delta
    \dot\Omega / \dot\Omega} \over {\Delta \Omega / \Omega } } \, { {|\Delta I
    |}\over{I_0} } \; = \; { {\Delta \dot\Omega / \dot\Omega} \over {\Delta
    \Omega / \Omega } } \; f \; {\icrust\over {I_0} } \; > \; (10^{-1}\; {\rm
  to} \; 10) \; f \; ,
\label{eq:omdot}
\eeqn where use of Eq.\ (2) has been made. Equation (3) yields a small $f$
value, i.e., $f < (10^{-4} \; {\rm to} \; 10^{-1})$, in agreement with $f\lsim
10^{-1}$ established just above. Here measured values of the ratio $(\Delta
\Omega/\Omega)/(\Delta\dot\Omega/\dot\Omega) \sim 10^{-6}$ to $10^{-4}$ for the
Crab and Vela pulsars, respectively, have been used.  So we arrive at the
important finding that the nuclear crust mass that can envelope a strange
matter core can be sufficiently large enough such that the relative changes in
$\Omega$ and $\dot\Omega$ obtained for strange stars with crust in the
framework of the crust quake model are consistent with the observed values, in
contrast to claims expressed in the literature.

\section{Cooling behavior of neutron stars and strange stars}

The left panel of Fig.\ 7 shows a numerical simulation of the thermal evolution
of neutron stars.  The neutrino emission rates are determined by the modified
and direct Urca processes, and the presence of a pion or kaon condensate.  The
baryons are treated as superfluid particles. Hence the neutrino emissivities
are suppressed by an exponential factor of $\exp(-\Delta/kT)$, where $\Delta$
is the width of the superfluid gap (see Schaab, Weber, Weigel, and
\begin{figure}[tb]
\begin{center}
\leavevmode
\mbox{\psfig{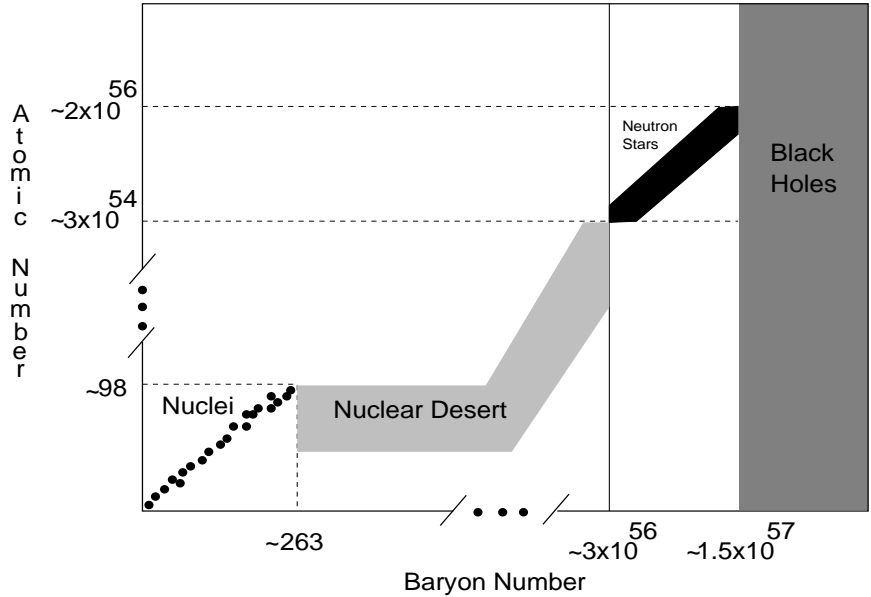}}
\caption[Graphical illustration of all possible stable nuclear objects if 
nuclear matter (i.e., iron) is the most stable form of matter.]  {\em Graphical
  illustration of all possible stable nuclear objects if nuclear matter (i.e.,
  iron) is the most stable form of matter. Note the huge range referred to as
  nuclear desert which is void of any stable nuclear systems.}
\end{center}
\end{figure}
\begin{figure}[tb]
\begin{center}
\leavevmode
\mbox{\psfig{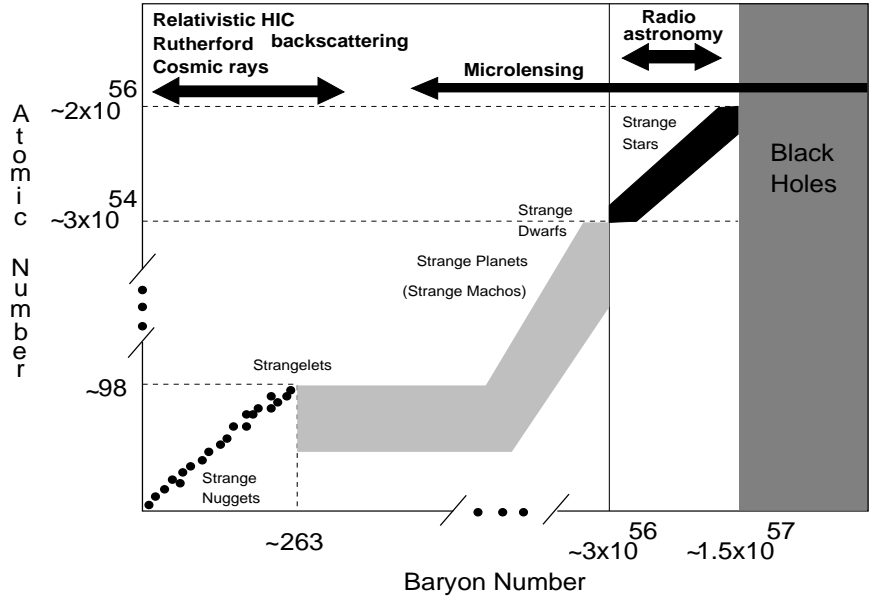}}
\caption[Graphical illustration of all possible stable nuclear objects if 
strange quark matter is more stable than nuclear matter.]  {\em Same as Fig.\ 8
  but for strange quark matter as the most stable configuration of matter.
  Various stable strange-matter objects are shown.  In sharp contrast to Fig.\ 
  8, the nuclear desert does not exist anymore but is filled with a variety of
  different stable strange-matter objects, ranging from strangelets at the
  small baryon number end to strange dwarfs at the large baryon number end. The
  strange counterparts of ordinary atomic nuclei are denoted strange nuggets,
  those of neutron stars (pulsars) are referred to as compact strange stars
  (see text for details).  Observational implications are indicated at the
  top.}
\end{center}
\end{figure} Glendenning\ (1996) for details).  Due to the dependence of the
direct Urca process and the onset of meson condensation on star mass, stars
that are too light for these processes to occur (i.e., $M<1\,\msun$) are
restricted to standard cooling via modified Urca.  Enhanced cooling via the
other three processes results in a sudden drop of the star's surface
temperature after about 10 to $10^3$ years after birth, depending on the
thickness of the ionic crust.  As one sees, agreement with the observed data is
achieved only if different masses for the underlying pulsars are assumed.  The
right panel of Fig.\ 7 shows cooling simulations of strange quark stars. The
curves differ with respect to assumptions made about a possible superfluid
behavior of the quarks. Because of the higher neutrino emission rate in
non-superfluid quark matter, such quark stars cool most rapidly (as long as
cooling is core dominated). In this case one does not get agreement with most
of the observed pulsar data. The only exception is pulsar PSR 1929+10.
Superfluidity among the quarks reduces the neutrino emission rate, which delays
cooling (Schaab, Weber, Weigel, and Glendenning,\ 1996). This moves the cooling
curves into the region where most of the observed data lie.

Subject to the inherent uncertainties in the behavior of strange quark matter
as well as superdense nuclear matter, at present it appears much too premature
to draw any definitive conclusions about the true nature of observed pulsars.
Nevertheless, should a continued future analysis in fact confirm a considerably
faster cooling of strange stars relative to neutron stars, this would provide a
definitive signature (together with rapid rotation) for the identification of a
strange star. Specifically, the prompt drop in temperature at the very early
stages of a pulsar, say within the first 10 to 50 years after its formation,
could offer a good signature of strange stars (Pizzochero, 1991).  This
feature, provided it withstands a more rigorous analysis of the microscopic
properties of quark matter, could become particularly interesting if continued
observation of SN 1987A would reveal the temperature of the possibly existing
pulsar at its center.

\section{Summary}

This work deals with an investigation of the properties of the complete
sequences of strange-matter stars that carry nuclear crusts.  Some striking
features of such objects are summarized in Table\ 3.  Figures 8 and 9 stress
the implications of strange quark matter as the most stable form of matter
graphically.  The following items are particularly noteworthy:

\begin{enumerate}
\item The complete sequence of compact strange stars can sustain
  extremely rapid rotation and not just those close to the mass peak,
  as is the case for neutron stars!

\item If the \smh is correct, the observed white dwarfs 
  and planets could contain strange-matter cores in their centers. The
  baryon numbers of their cores are smaller than $\lsim 2 \times
  10^{55}$!

\item The strange stellar configurations would populate a vast region in the
  mass-radius plane of collapsed stars that is entirely void of stars if
  strange quark matter is not the absolute ground state of strongly interacting
  matter!

\item If the new classes of stars mentioned in (2) and (3) exist
  abundantly enough in our Galaxy, the presently performed
  gravitational microlensing experiments could see them all!

\item We find that the moment of inertia of the crust on a strange star can
  account for both the observed relative frequency changes of pulsars
  (glitches) as well as the relative change in spin-down rate!

\item Due to the uncertainties in the behavior of superdense
  nuclear as well as strange matter, no definitive conclusions about
  the true nature (strange or conventional) of observed pulsar can be
  drawn from cooling simulations yet.  As of yet they could be made of
  strange quark matter as well as of conventional nuclear matter.
\end{enumerate}

Of course, there remain various interesting aspects of strange pulsars, strange
dwarfs and strange planets that need to be worked out in detail.  From their
analysis one may hope to arrive at definitive conclusions about the behavior of
superdense nuclear matter and, specifically, the true ground state of strongly
interacting matter. Clarifying the latter item is of fundamental importance
for the early universe, its evolution to the present day, massive stars,
and laboratory physics.

\medskip
{\bf Acknowledgment:}
\doe

\section{References}

\refind Alcock, C., Farhi, E., Olinto, A. V.: 1986, Astrophys.\ J.\ {\bf 310},
     p.\ 261.

\refind Alcock, C., Olinto, A. V.: 1988, Ann.\ Rev.\ Nucl.\ Part.\ Sci.\ 
     {\bf 38}, p.\ 161.

\refind Baym, G., Pethick, C., Sutherland, P.: 1971, Astrophys.\ J.\ {\bf 170},
     p.\ 299.

\refind Baym, G, Kolb, E. W., McLerran, L., Walker, T. P., Jaffe, R. L.:
     1985, Phys.\ Lett.\ {\bf 160B}, p.\ 181.

\refind Benvenuto, O. G., Horvath,  J. E.: 1989, Phys.\ Rev.\ Lett.\ {\bf 63},
     p.\ 716.

\refind Bjorken, J. D., McLerran, L.: 1979, Phys.\ Rev.\ D {\bf 20}, p.\ 2353.

\refind Br{\"{u}}gger, M., L{\"{u}}tzenkirchen, K., Polikanov, S., Herrmann, G.,
  Overbeck, M., Trautmann, N., Breskin, A., Chechik, R., Fraenkel, Z., 
  Smilansky, U.: 1989, Nature {\bf 337}, p.\ 434.

\refind Chin, S. A., Kerman, A. K.: 1979, Phys.\ Rev.\ Lett.\ {\bf 43}, p.\ 1292.

\refind De R{\'{u}}jula, A., Glashow, S. L., Wilson, R. R., Charpak, G.: 1983,
     Phys.\ Rep.\ {\bf 99}, p.\ 341.

\refind De R{\'{u}}jula, A., Glashow, S. L.: 1984, Nature {\bf 312}, p.\ 734.

\refind Farhi, E., Jaffe, R. L.: 1984, Phys.\ Rev.\ D {\bf 30}, p.\ 2379.

\refind Freedman, B. A., McLerran, L. D.: 1977, Phys.\ Rev.\ D {\bf 16}, 
     p.\ 1130; {\bf 16}, p.\ 1147; {\bf 16}, p.\ 1169.

\refind Frieman, J. A., Olinto, A. V.: 1989, Nature {\bf 341}, p.\ 633.

\refind Glendenning, N. K.: 1985, Astrophys.\ J.\ {\bf 293}, p.\ 470.

\refind Glendenning, N. K.: 1990, Mod.\ Phys.\ Lett.\ {\bf A5}, p.\ 2197.

\refind Glendenning, N. K., Weber, F.: 1992, Astrophys.\ J.\ {\bf 400}, p.\ 647.

\refind Glendenning, N. K., Weber, F.: 1994, Phys.\ Rev.\ D {\bf 50}, p.\ 3836.

\refind Glendenning, N. K., Kettner, Ch., Weber, F.: 1995a, Astrophys.\ J.\ 
     {\bf 450}, p.\ 253.

\refind Glendenning, N. K., Kettner, Ch., Weber, F.: 1995b, Phys.\ Rev.\ Lett.\ 
     {\bf 74}, p.\ 3519.

\refind Greiner, C., Koch, P., St{\"{o}}cker, H.: 1987, Phys.\ Rev.\ Lett.\ 
     {\bf 58}, p.\ 1825.

\refind Greiner, C., Rischke, D.-H., St{\"{o}}cker, H., Koch, P.: 1988,
     Phys.\ Rev.\ D {\bf 38}, p.\ 2797.

\refind Haensel, P., Zdunik, J. L., Schaeffer, R.: 1986, Astron.\ Astrophys.\ 
     {\bf 160}, p.\ 121.

\refind Horvath, J. E., Benvenuto, O. G.: 1988, Phys.\ Lett.\ {\bf 213B}, p.\
     516.

\refind Horvath, J. E., Benvenuto, O. G., Vucetich, H.: 1992,  Phys.\ Rev.\ D 
     {\bf 45}, p.\ 3865.

\refind Horvath, J. E., Vucetich, H., Benvenuto, O. G.: 1993, Mon.\ Not.\ R.\ 
     Astr.\ Soc.\ {\bf 262}, p.\ 506.

\refind Jaffe, R. L.: 1977, Phys.\ Lett.\ {\bf 38}, p.\ 195.

\refind Kettner, Ch., Weber, F., Weigel, M. K., Glendenning, N. K.: 1995a, 
     Phys.\ Rev.\ D {\bf 51}, p.\ 1440.

\refind Kettner, Ch., Weber, F., Weigel, M. K., Glendenning, N. K.: 1995b,
  Proceedings of the International Symposium on Strangeness and Quark Matter,
  eds. G. Vassiliadis, A. D. Panagiotou, S. Kumar, and J. Madsen, World 
  Scientific, p.\ 333.

\refind Liu, H.-C., Shaw, G. L.: 1984, Phys.\ Rev.\ D {\bf 30}, p.\ 1137.

\refind MACRO collaboration: 1992, Phys.\ Rev.\ Lett.\ {\bf 69}, p.\ 1860.

\refind Madsen, J., Heiselberg, H., Riisager, K.: 1986, Phys.\ Rev.\ D {\bf 34},
     p.\ 2947.

\refind Madsen, J.: 1988, Phys.\ Rev.\ Lett.\ {\bf 61}, p.\ 2909.

\refind Olinto, A. V.: 1987, Phys.\ Lett.\ {\bf 192B}, p.\ 71.

\refind Pizzochero, P.: 1991, Phys.\ Rev.\ Lett.\ {\bf 66}, p.\ 2425.

\refind Price, P. B.: 1984, Phys.\ Rev.\ Lett.\ {\bf 52}, p.\ 1265.

\refind Saito, T., Hatano, Y., Fukuda, Y., Oda, H.: 1990, Phys.\ Rev.\ Lett.\ 
     {\bf 65}, p.\ 2094.

\refind Saito, T.: 1995, Proceedings of the International Symposium on Strangeness
     and Quark, eds. G. Vassiliadis, A. D. Panagiotou, S. Kumar, and J. 
     Madsen, World Scientific.

\refind Schaab, Ch., Weber, F., Weigel, M. K., Glendenning, N. K.: 1996, Nucl.\ 
     Phys.\ {\bf A605}, p.\ 531.

\refind Shaw, G. L., Shin, M., Dalitz, R. H., Desai, M.: 1989, Nature {\bf 337},
     p.\ 436.

\refind Terazawa, H.: 1991, J.\ Phys.\ Soc.\ Japan, {\bf 60}, p.\ 1848.

\refind Terazawa, H.: 1993, J.\ Phys.\ Soc.\ Japan, {\bf 62}, p.\ 1415.

\refind Thomas, J., Jacobs, P.: 1995, {\it A Guide to the High Energy Heavy Ion
  Experiments}, UCRL-ID-119181.

\refind Weber, F., Weigel, M. K.: 1989, Nucl.\ Phys.\ {\bf A505}, p.\ 779.

\refind Weber, F., Glendenning, N. K.: 1992, Astrophys.\ J.\ {\bf 390}, p.\ 541.

\refind Weber, F., Glendenning, N. K.: 1993a, Proceedings of the Nankai Summer
  School, ``Astrophysics and Neutrino Physics'', ed. by D. H.  Feng, G. Z. He,
  and X. Q. Li, World Scientific, Singapore, p.\ 64--183.

\refind Weber, F., Glendenning, N. K.: 1993b, Proceedings of the First Symposium
     on Nuclear Physics in the Universe, ed. by M. W. Guidry and M. R. 
     Strayer, IOP Publishing Ltd, Bristol, UK, p.\ 127.

\refind Witten, E.: 1984, Phys.\ Rev.\ D {\bf 30}, p.\ 272.

\end{document}